\documentclass[12pt,twoside]{article}
\usepackage[utf8]{inputenc}

\usepackage{amsfonts}
\usepackage{amsmath,amssymb}
\usepackage{amsthm}
\usepackage{graphicx}
\usepackage{cite}
\usepackage{epsfig}
\usepackage{a4}
\usepackage{booktabs}

\newcommand{\I}{\mathrm{i}}        
\newcommand{\E}{\mathrm{e}}        
\newcommand{\D}{\mathrm{d}}       

\newcommand{\res}{\operatorname{Res}} 
\renewcommand{\Im}{\operatorname{Im}} 

\newcommand{\Det}{\operatorname{det}}

\newcommand{\End}{\operatorname{End}}   

\newcommand{\tr}{\operatorname{tr}} 

\newcommand{\ch}{\operatorname{ch}}  
\newcommand{\sh}{\operatorname{sh}}

\newcommand{\As}{\mathcal{A}}
\newcommand{\Bs}{\mathcal{B}}
\newcommand{\Cs}{\mathcal{C}}
\newcommand{\Ds}{\mathcal{D}}
\newcommand{\Ts}{\mathcal{T}}
\newcommand{\Ss}{\mathcal{S}}

\newcommand{\fraka}{\mathfrak{a}}
\newcommand{\frakb}{\mathfrak{b}}

\newcommand{\XXZ}{X\!X\!Z}
\newcommand{\XXX}{X\!X\!X}
\newcommand{\boundary}{\xi} 
\newcommand{\lain}{\zeta} 
\newcommand{\etal}{{\it et al.}}

\theoremstyle{plain}
\newtheorem*{example*}{Example}
\newtheorem*{conjecture*}{Conjecture}
\newtheorem*{proposition*}{Proposition}
\newtheorem*{propositionKKMNSTb*}{Proposition\cite{KKMNST07b}}
\newtheorem*{propositionKKMNSTa*}{Proposition\cite{KKMNST08a}}

\newtheorem*{lemma*}{Lemma}
\newtheorem{lemma}{Lemma}
\newtheorem*{remark}{Remark}

\theoremstyle{definition}

\newlength{\HFPP}       \HFPP5.4mm
\makeatletter
\@addtoreset{equation}{section}
\makeatother

\begin{document}

\thispagestyle{empty}

\begin{center}

{\Large {\bf Non-linear Integral Equations and Determinant Formulae of the Open XXZ Spin Chain\\}}

\vspace{7mm}

{\large Alexander Seel\footnote[1]{e-mail: alexander.seel@itp.uni-hannover.de} and Tobias Wirth\footnote[2]{e-mail: tobias.wirth@itp.uni-hannover.de} }

\vspace{5mm}

Institut f\"ur Theoretische Physik, Leibniz Universit\"at Hannover,\\
Appelstr. 2, 30167 Hannover, Germany\\[2ex]

\vspace{20mm}

{\large {\bf Abstract}}

\end{center}

\begin{quote}
We derive a non-linear integral equation for the Bethe-ansatz solvable open $\XXZ$ spin chain of arbitrary length describing the lowest lying state with zero magnetization. For this case we show how to combine the integral representation with the known determinant formula of norms and scalar products.\\[2ex] 

{\it PACS: 02.30.Ik, 75.10.Pq}
\end{quote}

\clearpage

%
\section{Introduction}
The doping of spin chains has directly observable consequences in experiments. For example the magnetic susceptibility of a spin-$1/2$ Heisenberg chain made of $Sr_2$$CuO_3$ shows a strong dependency upon the oxygen content \cite{ACHWJHE95}. In one spatial dimension the impurities cut the chain and serve as effective boundary fields. This breaks the translational invariance and makes the local magnetization dependent on the lattice site. However, by introducing an additional reflection algebra Cherednik \cite{Cherednik84} and Sklyanin \cite{Sklyanin88} showed the open Heisenberg chain to remain integrable. For diagonal boundaries,
\begin{align} \notag
\mathcal{H} = \sum_{j=1}^{L-1}&\Big[\sigma_j^x\sigma_{j+1}^x + \sigma_j^y\sigma_{j+1}^y +\big(\sigma_j^z\sigma_{j+1}^z + 1\big)\ch\eta\Big] +\ch\eta\\ \label{hamil}
+&\Big[\sigma_1^z\coth\boundary^- + \sigma_L^z\coth\boundary^+\Big]\sh\eta \quad ,
\end{align}
it can even be solved by the coordinate \cite{ABBBQ87} or the algebraic Bethe ansatz \cite{Sklyanin88}. $L$ is the number of lattice sites and $\sigma_j^\alpha$, $\alpha=x,y,z$ are the usual Pauli matrices. The boundary fields are parametrized by the complex numbers $\xi^\pm$ and $\eta$ is the crossing parameter of the model entering the anisotropy $\Delta=\ch\eta$. The distribution of Bethe roots in the ground state depends on the boundary fields. For example anisotropies $|\Delta|<1$ allow for at most two purely imaginary Bethe numbers besides real roots \cite{SkSa95}. From the coordinate Bethe ansatz such states containing imaginary rapidities are usually termed \lq boundary bound states\rq. This terminology is related to the exponential decay of phase factors. The ground state for anisotropies $\Delta>1$ can be found in \cite{KaSk96}.

To calculate the $S^z$-magnetization in the ground state as an expectation value in the framework of the algebraic Bethe ansatz one could make use of the inverse problem Wang \cite{Wang00} solved in terms of a mixture of the reflection and the Yang-Baxter algebra. However, this method makes use of the translation operator for which the Bethe states are no longer eigenstates. This difficulty Kitanine \etal\ overcame by reducing the problem to the algebra of the periodic chain where the inverse problem \cite{KMT99a} is only expressed in terms of this algebra. So they determined the action of local operators on Bethe vectors in the representation of the reflection algebra and thus were able to apply the trigonometric generalization \cite{KKMNST07b} of the rational determinant formula \cite{Wang02} for scalar products. Additionally they succeeded in simplifying the combinatorial part of the local $S^z$-magnetization by introducing a generating function \cite{KKMNST08a}. The result was a multiple integral representation for its state average value. The integral representations is linked to expressions from the vertex operator approach \cite{JKKKM95} and was derived for the ground state of \eqref{hamil} described by Bethe root densities which are valid in the thermodynamic limit $L\to\infty$ of the half infinite chain. 

Treating finite chains arises from the subtle observation that on a formal level integrable systems of finite length share much of the properties of systems at finite temperature: In the first case the logarithmic derivative of the eigenvalue of the usual row-to-row transfer matrix \cite{Baxter82} determines the ground state energy of the system of length $L$. In the second case the free energy in the thermodynamic limit is connected to the leading eigenvalue of the quantum transfer matrix \cite{Suzuki85,SuIn87} at temperature $T$. As recently shown by Damerau \etal \cite{DGHK07}, utilizing this analogy the method of non-linear integral equations \cite{KlBa90,KlBaPe91} describing the quantum transfer matrix could easily be applied to the usual row-to-row transfer matrix of the finite-length periodic Heisenberg chain to calculate high-precision data of correlation functions. Extending this approach to the reflection algebra of the open $\XXZ$ spin chain will be the scope of this article.\\

The paper is organized as follows. In section \ref{IBC} we will review the reflection algebra. Then in section \ref{auxfunc} we give a non-linear integral equation for an auxiliary function
accounting for the lowest lying state of \eqref{hamil} with zero magnetization. In section \ref{Detplusaux} we show our main result, how to combine the known determinant formula with the auxiliary function to represent normalized scalar products in terms of multiple integrals. Section \ref{genfunkmag} is devoted to an example.


\section{Integrable Boundary Conditions} \label{IBC}
Sklyanin's construction \cite{Sklyanin88} of integrable systems involving
boundaries is valid for a general class of integrable systems characterized by
an $R$-matrix of difference form $R(\lambda,\mu) =$ \mbox{$R(\lambda-\mu)$}$
\in \End(V\otimes V)$ ($V$ is a vector space with $\dim V  \in \mathbb{N}$ )
which not only satisfies the Yang-Baxter equation 
\begin{equation} \label{YBE}
R_{12}(\lambda-\mu)\,R_{13}(\lambda-\nu)\,R_{23}(\mu-\nu)
= R_{23}(\mu-\nu)\,R_{13}(\lambda-\nu)\,R_{12}(\lambda-\mu)
\end{equation}
but also several conditions such as symmetry with respect to the permutation
operator $P$ on $V \otimes V$ ($P\,x\otimes y = y \otimes x$), 
\begin{equation}
R(\lambda) = PR(\lambda)P\quad ,
\end{equation}
unitarity involving some complex function $\rho(\lambda)$,
\begin{equation}
R(\lambda)R(-\lambda) = \rho(\lambda)
\end{equation}
and crossing unitarity for another complex function $\widetilde\rho(\lambda)$,
\begin{equation}
R^{t_1}(\lambda)R^{t_1}(-\lambda-2\eta) = \widetilde\rho(\lambda) \quad.
\end{equation}
The parameter $\eta$ characterizes the $R$-matrix and the superscript $t_j$
denotes the transposition with respect to the $j$th space in the tensor
product $V\otimes V$. Here we will need the well-known $6$-vertex model
solution  
\begin{equation}\label{Rmatrix}
R(\lambda) =
  \begin{pmatrix}
     \sh(\lambda+\eta) & 0 & 0 & 0 \\
     0 & \sh\lambda & \sh\eta & 0 \\
     0 & \sh\eta & \sh\lambda & 0 \\
     0 & 0 & 0 & \sh(\lambda+\eta)
  \end{pmatrix}
\end{equation}
of the Yang-Baxter equation \eqref{YBE}. It generates the Hamiltonian of the
spin-$\frac12$ $\XXZ$ chain with 
\begin{equation}
\rho(\lambda) = \sh(\lambda+\eta)\,\sh(-\lambda+\eta) \quad ,\quad \widetilde\rho(\lambda) = {\sh(-\lambda)\,\sh(\lambda+2\eta)} \quad .
\end{equation}
Each solution $R(\lambda)$ of the Yang-Baxter equation fixes the structure
constants of a Yang-Baxter algebra  
\begin{equation}\label{YBA}
R_{12}(\lambda-\mu) T_1(\lambda) T_2(\mu) = T_2(\mu) T_1(\lambda)R_{12}(\lambda-\mu)
\end{equation}
with generators $T^\alpha_{\phantom{x}\beta}(\lambda)$, $\alpha,\beta=1,2$;
where $T_1(\lambda) = T(\lambda) \otimes I_2$, $T_2(\lambda) =I_2 \otimes
T(\lambda)$ are the embeddings of the monodromy matrix $T(\lambda)$ with $2\times2$ unity $I_2$. \\

Sklyanin's construction of open spin chains is based on the representations of
two algebras $\Ts^{(+)}$ and $\Ts^{(-)}$ defined by the relations
\begin{equation}\label{leftalg}
R_{12}(\lambda-\mu) \Ts_1^{(-)}(\lambda) R_{12}(\lambda+\mu)\Ts_2^{(-)}(\mu) 
=\Ts_2^{(-)}(\mu)R_{12}(\lambda+\mu)\Ts_1^{(-)}(\lambda)R_{12}(\lambda-\mu)
\end{equation}
\begin{multline}\label{rightalg}
R_{12}(-\lambda+\mu) \Ts_1^{(+)t_1}(\lambda) R_{12}(-\lambda-\mu-2\eta)\Ts_2^{(+)t_2}(\mu) =\\
=\Ts_2^{(+)t_2}(\mu)R_{12}(-\lambda-\mu-2\eta)\Ts_1^{(+)t_1}(\lambda)R_{12}(-\lambda+\mu) \quad  .
\end{multline}
We shall call $\Ts^{(+)}$ and $\Ts^{(-)}$ right and left reflection algebras
respectively. The transfer matrix  
\begin{equation}
t(\lambda) = \tr \Ts^{(+)}(\lambda)\Ts^{(-)}(\lambda)
\end{equation}
as a trace in auxiliary space is the central object under consideration
because it generates with $[t(\lambda),t(\mu)]=0$ a commuting family of
operators.
The explicit construction of integrable open boundary conditions for models
arising from the Yang-Baxter algebra starts with the $2\times2$ matrix  
\begin{equation} \label{kmatrix}
K(\lambda,\boundary) = \frac{1}{\sh\xi \ch\lambda}
\begin{pmatrix}
\sh(\lambda+\boundary) &  \\
 & -\sh(\lambda-\boundary)
\end{pmatrix} 
  =I_2 + \sigma^z \,\tanh\lambda \,\coth\boundary 
\end{equation}
originally found by Cherednik \cite{Cherednik84} and cast into the form \eqref{kmatrix} by de~Vega \etal \cite{VeGo93}.
It constitutes the $c$-number representations $K^{(+)}(\lambda) =
\frac12 K(\lambda+\eta,\boundary^+)$ and $K^{(-)}(\lambda) = K(\lambda,\boundary^-)$ of the
reflection algebras with the obvious properties
\begin{equation}
\tr K(\lambda,\boundary) = 2\,, \quad 
K^{(-)}(0) = I_2\,, \quad  \tr K^{(+)}(0) = 1 \,.
\end{equation}

The Hamiltonian \eqref{hamil} involving the interaction of two neighbouring sites is connected, up to a factor, to the first derivative of
$t(\lambda)$ by looking at the expansion $t(\lambda) = 1 +
2\lambda\mathcal{H}+\ldots$ around the point $\lambda=0$.  Considering local
$L$-matrices building up the two representations
$T^{(+)}(\lambda)=L_L(\lambda)\cdots L_{M+1}(\lambda)$ and
$T^{(-)}(\lambda)=L_M(\lambda)\cdots L_1(\lambda)$ of \eqref{YBA} then by
construction
\begin{equation}
\begin{aligned}
\Ts^{(-)}(\lambda) &= T^{(-)}(\lambda) K^{(-)}(\lambda) T^{(-)-1}(-\lambda) \\
\Ts^{(+)t}(\lambda) &= T^{(+)t}(\lambda) K^{(+)t}(\lambda) \big(T^{(+)-1}\big)^t(-\lambda)
\end{aligned}
\end{equation}
are representations of the reflection algebras such that the normalized transfer matrix
\begin{equation}
t(\lambda) 
= \tr K^{(+)}(\lambda) T(\lambda) K^{(-)}(\lambda)T^{-1}(-\lambda) \quad , \quad t(0)=1
\end{equation}
is independent of the factorization of $T(\lambda)=T^{(+)}(\lambda)T^{(-)}(\lambda)$. Thus we are free to choose
\begin{equation}
\Ts^{(+)t}(\lambda)= T^{t}(\lambda) K^{(+)t}(\lambda) \big(T^{-1}\big)^t(-\lambda)\quad , \quad 
\Ts^{(-)}(\lambda) =  K^{(-)}(\lambda) \quad .
\end{equation}
In order to gain more symmetric arguments and to avoid inconvenient scalar
functions after applying the inversion formula 
\begin{equation}
T^{-1}(\lambda) = \frac{1}{(d_qT)(\lambda-\eta/2)} \sigma^y T^t(\lambda-\eta) \sigma^y
\end{equation} 
it is instructive to define the new object 
$U^{(+)}(\lambda+\eta/2):=\Ts^{(+)}(\lambda)\,(d_q T)(-\lambda-\eta/2)$
consisting of
\begin{equation}\label{defU}
U^{(+)t}(\lambda)=T^t(\lambda-\eta/2)K^{(+)t}(\lambda-\eta/2) \sigma^y T(-\lambda-\eta/2)\sigma^y 
\quad .
\end{equation}
It is still a representation of the right reflection algebra with a $2\times2$
matrix in auxiliary space, 
\begin{equation}
U^{(+)}(\lambda) = 
\begin{pmatrix}
\As(\lambda) & \Bs(\lambda) \\
\Cs(\lambda) & \Ds(\lambda)
\end{pmatrix} \quad .
\end{equation}

The quantum determinant $(d_q T)(\lambda)$ is the central element (Casimir) of
the Yang-Baxter algebra (\ref{YBA}).  With the one-dimensional projector
$P_{12}^-$ onto the antisymmetric (singlet) state in the tensor product
$V\otimes V$ of auxiliary spaces the definition reads
\begin{equation}
\begin{aligned}
(d_q T)(\lambda) &= \tr_{12} P_{12}^- T_1(\lambda-\eta/2)T_2(\lambda+\eta/2) \\[.2\baselineskip]
 &= A(\lambda+\eta/2)\,D(\lambda-\eta/2) - B(\lambda+\eta/2)\,C(\lambda-\eta/2) \quad .
\end{aligned}
\end{equation}
Here, the trace $\tr_{12}$ is to be taken in both auxiliary spaces $1$ and $2$
of the tensor product $V\otimes V$ and the monodromy matrix $T$ enters with
the usual representation
\begin{equation}
T(\lambda) = 
\begin{pmatrix}
A(\lambda) & B(\lambda) \\
C(\lambda) & D(\lambda)
\end{pmatrix} \quad .
\end{equation}

As an example, implying \eqref{defU} the operator $\Bs(\lambda)$ can be reduced to the operators of the periodic chain via
\begin{multline} \label{Bop}
\Bs(\lambda)= \frac{1}{2\ch(\lambda+\eta/2)\sh\boundary^+}
\frac{\sh(2\lambda+\eta)}{\sh(2\lambda)}
\Big[\sh(\lambda-\eta/2+\boundary^+)B(\lambda-\eta/2)D(-\lambda-\eta/2) \\
+ \sh(\lambda+\eta/2-\boundary^+)B(-\lambda-\eta/2)D(\lambda-\eta/2)\Big]  \quad .
\end{multline}

\begin{remark}
The related transfer matrix
\begin{equation}  
\tau(\lambda)=\tr U^{(+)}(\lambda)K^{(-)}(\lambda-\eta/2)=t(\lambda-\eta/2)(d_qT)(-\lambda)
\end{equation}
to the monodromy $U^{(+)}(\lambda)$ reduces at the point $\lambda=\eta/2$ to $(d_qT)(-\eta/2)$. As the Hamiltonian is proportional to $t^\prime(0)$ it is now related to the logarithmic derivative of $\tau(\lambda)$ at $\lambda=\eta/2$ because of
$t^\prime(0)=\partial_\lambda\ln\tau(\eta/2)-\partial_\lambda\ln(d_qT)(-\eta/2)$.
\end{remark}


\section{Auxiliary Function} \label{auxfunc}
The open Heisenberg chain \eqref{hamil} is related to the fundamental representation
\begin{equation}
T(\lambda) = 
R_{0L}(\lambda-s_L)\ldots R_{02}(\lambda-s_2)R_{01}(\lambda-s_1)
\end{equation}
of the ordinary monodromy matrix
in auxiliary space $0$ for $L$ lattice sites each equipped with an inhomogeneity $s_j\in\mathbb{C}$. Choosing the vacuum $|0\rangle=\left(
\begin{smallmatrix}
1\\0
\end{smallmatrix}\right)^{\otimes L}$ for the algebraic Bethe ansatz the expectation values of the operators $A(\lambda)$ and $D(\lambda)$ read
\begin{equation}
a(\lambda)=\prod_{l=1}^L\sh(\lambda-s_l+\eta) \quad , \quad  d(\lambda)=\prod_{l=1}^L\sh(\lambda-s_l)
\end{equation} 
and $T(\lambda)$ acts as an upper triangular matrix. Because of \eqref{defU} the monodromy matrix $U^{(+)}(\lambda)$ of the right reflection algebra acts as an upper triangular matrix with respect to $|0\rangle$ as well enabling the algebraic Bethe ansatz \cite{Sklyanin88} with diagonal boundary fields $\boundary^\pm$. The quantum determinant takes the explicit form
\begin{equation}
(d_qT)(\lambda)=a(\lambda+\eta/2)d(\lambda-\eta/2)
\end{equation}
and for the new transfer matrix 
$\tau(\lambda)=\tr U^{(+)}(\lambda)K^{(-)}(\lambda-\eta/2)$ the corresponding Bethe ansatz equations \cite{Sklyanin88} for $M\leq L/2$ Bethe numbers are given by 
\begin{multline}\label{BAeq}
\frac{\sh(\lambda_j-\boundary^++\eta/2)\sh(\lambda_j-\boundary^-+\eta/2)}
{\sh(\lambda_j+\boundary^+-\eta/2)\sh(\lambda_j+\boundary^--\eta/2)}
\Bigg[\prod_{l=1}^L\frac{\sh(\lambda_j-\eta/2+s_l)\sh(\lambda_j-\eta/2-s_l)}
{\sh(\lambda_j+\eta/2+s_l)\sh(\lambda_j+\eta/2-s_l)}\Bigg] =\\
=\bigg[\prod_{\overset{\scriptstyle{l=1}}{l\not=j}}^M
\frac{\sh(\lambda_j-\lambda_l-\eta)\sh(\lambda_j+\lambda_l-\eta)}
{\sh(\lambda_j-\lambda_l+\eta)\sh(\lambda_j+\lambda_l+\eta)}\bigg]
\end{multline}
for all $j=1,\ldots,M$ rendering the eigenvalue
\begin{multline} \label{eigenvalueexplicit}
\Lambda(z) = \frac{(-1)^L \, \phi(z+\eta/2)}{2\ch(z+\eta/2)\ch(z-\eta/2)}
\frac{\sh(2z+\eta)}{\sh(2z)}\frac{\sh(z+\boundary^+-\eta/2)}{\sh\boundary^+}
\frac{\sh(z+\boundary^--\eta/2)}{\sh\boundary^-}\frac{q(z-\eta)}{q(z)}\\
+\frac{(-1)^L \, \phi(z-\eta/2)}{2\ch(z+\eta/2)\ch(z-\eta/2)}
\frac{\sh(2z-\eta)}{\sh(2z)}\frac{\sh(z-\boundary^++\eta/2)}{\sh\boundary^+}
\frac{\sh(z-\boundary^-+\eta/2)}{\sh\boundary^-}\frac{q(z+\eta)}{q(z)}
\end{multline}
analytic at the Bethe roots $\lambda_j$ defining the function $q(z):=\big[\prod_{l=1}^M\sh(z-\lambda_l)\sh(z+\lambda_l)\big]$. The shorthand \mbox{$\phi(z):=\big[\prod_{l=1}^L\sh(z-s_l)\sh(z+s_l)\big]$} accounts for the pairwise distinct lattice inhomogeneities $s_j$ regularizing combinatorial expressions in the forthcoming sections.

\begin{remark}
Note that the Hamiltonian \eqref{hamil} corresponds to the homogeneous case $s_j\to0$ such that we have to perform this limit (including $\partial_\lambda\ln(d_qT)(-\eta/2)=0$) if we want to compare e.g.\ with results from the exact diagonalization.
\end{remark}

\begin{figure}
\begin{center}
\includegraphics[height=4.8cm]{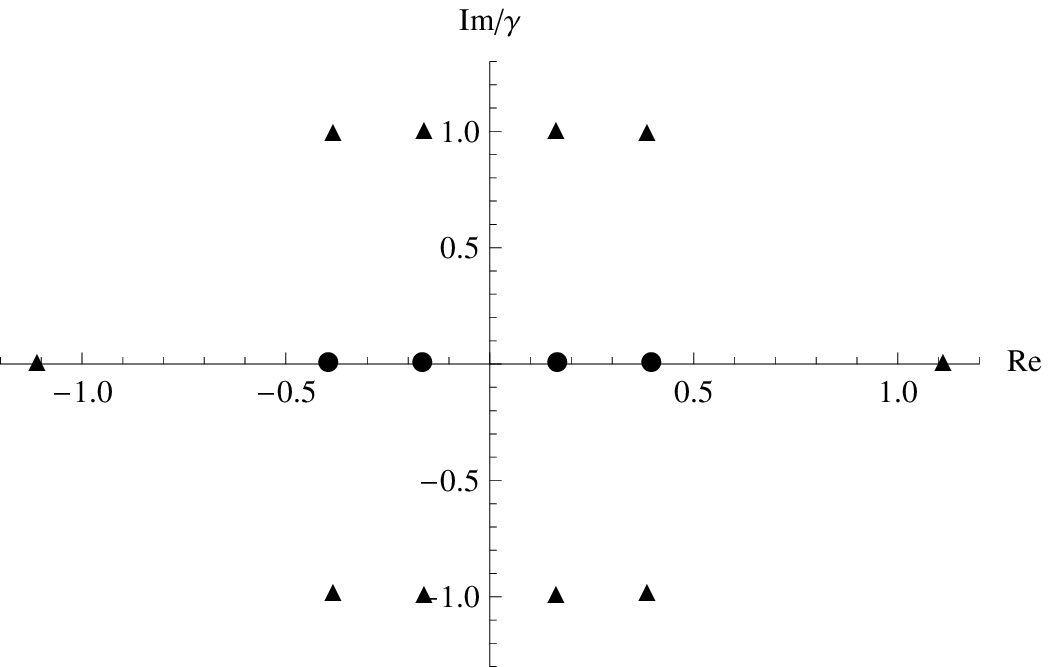} \hspace{1.5em}\includegraphics[height=4.8cm]{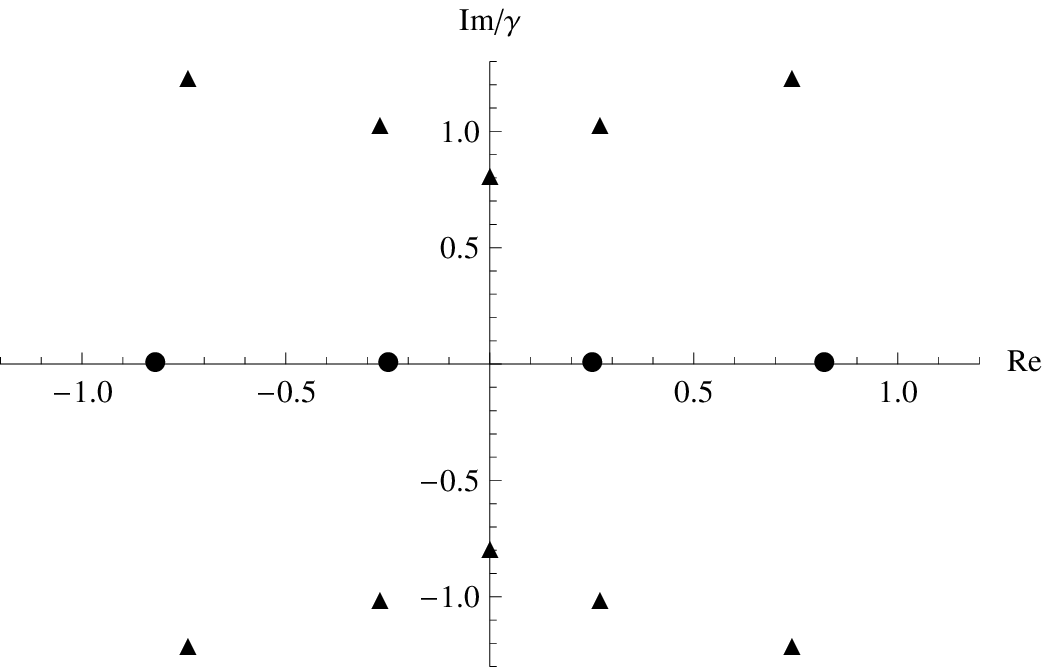}
\caption{\label{BAroot} Solution of $1+\fraka(z)=0$ in the rational limit for the ground state of $L=4$ lattice sites: A typical distribution of (two) Bethe roots $\bullet$ and (five) hole-type solutions $\blacktriangle$ for positive boundary fields $\boundary^+=1.1\I$, $\boundary^-=.9\I$ (left panel) and a negative boundary field $\boundary^-=-.3\I$ along with $\boundary^+=4\I$  (right panel) in the sector of $M=2$.}
\end{center}
\end{figure}

Let us restrict the anisotropy $\ch\eta$ of the $zz$-interaction to the massless case, $\eta=\I\gamma$, and choose $0 < \gamma < \pi/2$ for the region next to the isotropic point. By selecting the lowest lying state of zero magnetization, not necessarily the ground state, from Bethe vectors $\Bs(\lambda_1)\ldots\Bs(\lambda_M)|0\rangle$ with $M=L/2$ some simplifications occur for the auxiliary function 
\begin{multline} \label{auxfuncfactor}
\fraka(z):= \frac{\sh(z-\boundary^++\I\gamma/2)}{\sh(z+\boundary^+-\I\gamma/2)}
\frac{\sh(z-\boundary^-+\I\gamma/2)}{\sh(z+\boundary^--\I\gamma/2)}
\frac{\sh(2z-\I\gamma)}{\sh(2z+\I\gamma)}\\
\times\bigg[\prod_{l=1}^L\frac{\sh(z-\I\gamma/2+s_l)\sh(z-\I\gamma/2-s_l)}
{\sh(z+\I\gamma/2+s_l)\sh(z+\I\gamma/2-s_l)}\bigg]
\frac{q(z+\I\gamma)}{q(z-\I\gamma)}
\end{multline}
associated with the unique solution $\{\lambda_l\}_{l=1}^{L/2}=:\{\lambda\}$, $1+\fraka(\lambda_j)=0$. Obviously $L$ has to be even and both eigenvalue and auxiliary function are periodic in $\I\pi$. Because of periodicity the boundary parameters $\boundary^\pm$ can be restricted to the complex interval $(-\I\pi/2,\I\pi/2]$ for an hermitian Hamiltonian \eqref{hamil}. Once a set of Bethe numbers $\{\lambda\}=\{\lambda_l\}_{l=1}^{L/2}$ is fixed satisfying $1+\fraka(\lambda_j)=0$ for all $j=1,\ldots,L/2$ there are additional hole-type solutions $\{\chi\}=\{\chi_k\}_{k=1}^{L+1}$ to the same equation, $1+\fraka(\chi_k)=0$. These are also zeroes of the eigenvalue \eqref{eigenvalueexplicit},
\begin{equation*}
\Lambda(z) = \frac{(-1)^L \, \phi(z+\eta/2)\,q(z-\eta)}{2\ch(z+\eta/2)\ch(z-\eta/2)}
\frac{\sh(2z+\eta)}{\sh(2z)}\frac{\sh(z+\boundary^+-\eta/2)}{\sh\boundary^+}
\frac{\sh(z+\boundary^--\eta/2)}{\sh\boundary^-}\frac{1+\fraka(z)}{q(z)}\,\,.
\end{equation*}
The number of holes follow from the transformation $w:=\E^{2z}$ of $1+\fraka(z)$ into a rational function of $w$ where the nominator is a polynomial of degree $3L+4$: 
Due to the symmetry $\fraka(-z)=1/\fraka(z)$ all zeroes $\lambda_j$ and $\chi_k$ appear twice with different signs and thus they are symmetrically distributed with respect to the origin as shown in figure \ref{BAroot}. Additionally the equation $1+\fraka(z)=0$ has two trivial solutions $z=0$ and $z=\I\pi/2$ fixing the number of hole-type solutions to be $L+1$. \\

Compared to the case of the half infinite chain where the problem can be treated by root densities in the thermodynamic limit we want to pursue another way \cite{DGHK07,KlBa90,KlBaPe91} valid for a finite number of lattice sites. It turns out that the meromorphic function $\fraka(z)$ is sufficiently well determined by the gross properties of $\{\lambda\}$ and $\{\chi\}$ depending on the value of both boundary fields. As \eqref{auxfuncfactor} is symmetric in the parameters $\boundary^\pm$,
\begin{table}
\begin{center}
{\footnotesize
\begin{tabular}{ccccc}
\toprule[1.5pt]
&  $\pi/2 \geq \boundary^+/\I > \gamma$ &  $\gamma \geq \boundary^+/\I> \gamma/2$ &  $\gamma/2 \geq \boundary^+/\I>0$ &  $0 > \boundary^+/\I > -\pi/2$ \\
\midrule
 $\pi/2 \geq \boundary^-/\I > \gamma$ & $I$ &  &   & \\ 
 $\gamma \geq \boundary^-/\I> \gamma/2$ & $I\!I$ & $I\!I\!I$ && \\ 
 $\gamma/2 \geq \boundary^-/\I>0$ & $IV$ & $V$& $V\!I$& \\ 
 $0 > \boundary^-/\I> -\pi/2$ & $V\!I\!I$ & $V\!I\!I\!I$ & $I\!X$ & $X$\\
\bottomrule[1.5pt]
\end{tabular}}
\caption{\label{combinations}{Ten possible combinations of the boundary fields $\boundary^\pm$
}}
\end{center}
\end{table}
one has to distinguish between ten main cases, c.f.\ table \ref{combinations}, for the pole structure in view of $\boundary^\pm$, the positions of Bethe numbers and hole-type solutions. Indeed, the last case $X$ is sensitive to the values of $\boundary^\pm$ but inverting all parameters $\boundary^\pm\to-\boundary^\pm$ formally reverses the $z$-direction and maps the region $X$ to the cases $I$ to $V\!I$.\\

\begin{figure}
\begin{center}
\epsfig{file=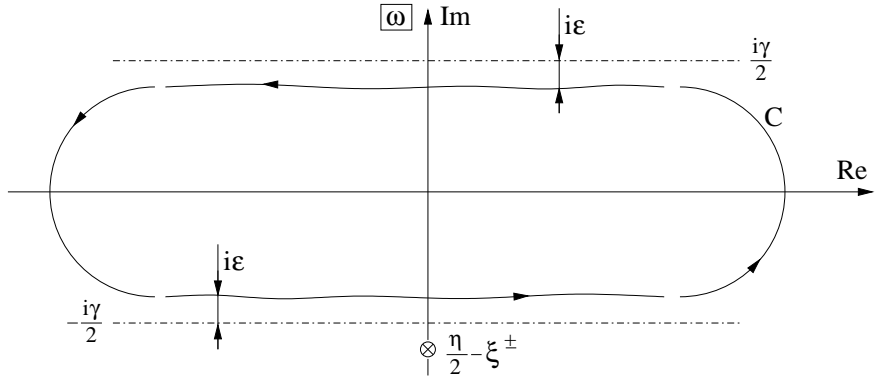,height=5cm}
\caption{\label{CKont} The Bethe roots are enclosed by the canonical contour $\mathcal{C}$ for the massless case in its parametrization $\eta=\I\gamma$, $0<\gamma<\pi/2$. Note the symmetry $\fraka(-z)=1/\fraka(z)$ mirroring all solutions of the equation $1+\fraka(z) = 0$ at the origin. The poles at $\omega=\eta/2-\boundary^\pm$ of the function $(1+\fraka)$ have to lie outside the contour.}
\end{center}
\end{figure}

By means of table \ref{combinations} lets consider some examples by looking at the first column. From numerical evidence one observes for opposite boundary fields (region $V\!I\!I$) $L/2$ real Bethe roots inside the strip $|\Im z|<\gamma/2$ and the hole-type solutions to lie outside of it. Additionally one hole-type solution seems to stick to the pole $z=\eta/2-\boundary^-$ of the boundary field (figure \ref{BAroot}, right panel). This observation along with the eigenvalue and the known asymptotics is enough to derive a set of equations relating the second logarithmic derivatives of $\fraka$ and $(1+\fraka)$ to each other determining $\fraka(z)$ uniquely by means of the integral Fourier transform. As this technique is explained in detail in \cite{KlBaPe91} we may leave with the homogeneous $s_j\to0$ result
\begin{align} 
\ln\fraka(z) &= 4\eta + \ln\Big[\frac{\sh(2z-\eta)\sh(z-\eta)}{\sh(2z+\eta)\sh(z+\eta)}\Big] \notag\\[.5ex]
&- 2\eta +\ln\Big[\frac{\sh(z-(\boundary^+-\eta/2))\sh(z-(\boundary^--\eta/2))}{\sh(z+(\boundary^+-\eta/2))\sh(z+(\boundary^--\eta/2))}\Big] \notag\\[.5ex]
&+ 2\eta L 
+ 2L \ln\Big[\frac{\sh(z-\eta/2)}{\sh(z+\eta/2)}\Big] - \int_{\mathcal{C}}\frac{\D\omega}{2\pi\I} \frac{\sh(2\eta)\ln(1+\fraka(\omega))}{\sh(z-\omega+\eta)\sh(z-\omega-\eta)} \label{NLIE}
\end{align}
valid for the region $|\Im z|\leq\gamma/2-\varepsilon$. The factor $\varepsilon\ll1$ ensures the Fourier integrals to converge and serves in the inhomogeneous case as a convenient restriction $|\Im s_j|<\varepsilon$. The canonical contour $\mathcal{C}$ is depicted in figure \ref{CKont} and extends to infinity. 
Compared to the periodic chain \cite{DGHK07} only the first and second lines of driving terms were added.

For boundary fields exceeding $\I\gamma$ (region $I$) a hole-type solution on the real line appears besides the outermost Bethe root (figure \ref{BAroot}, left panel). This is due to a change in the description of the ground state encoded by $L/2$ real Bethe roots. All other $L$ hole-type solutions remain outside the strip $|\Im z|<\gamma/2$. The hole-type solution on the real line corresponds to the term
\begin{equation}\label{drivinghole}
+ 4\eta + \ln\Big[\frac{\sh(z+\chi-\eta)\sh(z-\chi-\eta)}{\sh(z+\chi+\eta)\sh(z-\chi+\eta)}\Big]
\end{equation}
which has to be added on the RHS of \eqref{NLIE} due to complex analysis imposing the additional constraint $1+\fraka(\chi)=0$ on the auxiliary function. This is similar to the case of excited states in the periodic $\XXZ$ chain \cite{KRMSS00}. 

For boundary fields $\gamma-\varepsilon > \boundary^-/\I > 0$ (regions $I\!I$, $IV$) the pole at $z=\eta/2-\boundary^-$ has to remain outside the contour guaranteed by a deformation. Applying the residue theorem yields the additional driving term
\begin{equation} \label{driving}
-2\eta + \ln\Big[\frac{\sh(z+\eta/2+\boundary^-)}{\sh(z-3\eta/2+\boundary^-)}\Big]
\end{equation} 
which, besides \eqref{drivinghole}, has to be added on the RHS of (\ref{NLIE}) for that case. This is the only modification compared to case $I$ because the structure of the root distribution of $\lambda_j$ and $\chi_k$ with respect to the strip $|\Im z|<\gamma/2$ is unchanged. Nevertheless approaching with the pole $z=\eta/2-\boundary^-$ zero from below the Bethe root closest to the origin moves towards zero along the real axis (region $I\!I$). Passing the origin the pole picks up this Bethe root and pulls it 
(up to exponential corrections with respect to the chain length) 
along the positive imaginary axis until the upper part of the canonical contour is reached. Because of this imaginary Bethe root the corresponding states are termed \lq boundary bound states\rq\ (region $IV$).

Especially in the $\XXX$ limit when the parallely oriented boundary fields (region $V\!I$, two imaginary Bethe roots) become strong enough to significantly arrest the outermost spins of the chain the considered \lq boundary bound state\rq\ with $M=L/2$ Bethe roots and a total magnetization of zero refers no longer to the ground state. Here the almost fixed boundary spins can be regarded as effective boundary fields for a spin chain with two lattice sites and one Bethe root less. For the $\XXZ$ chain this effect already sets in for the regions $V$, $V\!I$, $V\!I\!I\!I$, $I\!X$ but depends on the values of the anisotropy $\ch\eta$ compared to the boundary fields.\\

\enlargethispage{.8\baselineskip}
Numerics suggest for all cases $I$ to $I\!X$ with zero magnetization to have $L/2$ Bethe numbers within the contour $\mathcal C$ as in the considered examples above. Thus all possible forms of driving terms with respect to hole-type solutions and poles of the boundary fields are given. The non-linear integral equation for the auxiliary function can then be fixed if one considers {table~\ref{drivterms}}
\begin{table}
\begin{center}
\begin{tabular}{c c c c c c c c c c}
\toprule[1.5pt]
&$I$ & $I\!I$ & $I\!I\!I$ & $IV$ & $V$ & $V\!I$ & $V\!I\!I$ & $V\!I\!I\!I$ & $I\!X$\\
\midrule
hole-type solution& $\bullet$ & $\bullet$ & $\bullet$ & $\bullet$ & $\bullet$ & $\bullet$&&&\\
left boundary pole $(-)$&&$\bullet$&$\bullet$&$\bullet$&$\bullet$&$\bullet$&&&\\
right boundary pole $(+)$&&&$\bullet$&&$\bullet$&$\bullet$&&$\bullet$&$\bullet$\\
\bottomrule[1.5pt]
\end{tabular}
\caption{\label{drivterms}Driving terms for the possible combinations of boundary fields $\boundary^\pm$}
\end{center}
\end{table}
as a building block. 
Here $\bullet$ marks the additional driving terms \eqref{drivinghole} and \eqref{driving} which have to be added on the RHS of \eqref{NLIE} for each single case. This accounts for the hole-type solution $\chi$ inside the canonical contour imposing $1+\fraka({\chi})=0$ and the boundary fields $\boundary^\pm$.\\
\enlargethispage{.5\baselineskip}

In the following two sections we shall derive our main result valid for distributions of Bethe numbers in the strip $|\Im z|<\gamma/2$ according to the left panel of figure \ref{BAroot}. For this reason, we have to introduce a closed contour $\mathcal{C^\prime}$ similar to $\mathcal{C}$ but excluding all hole-type solutions, especially the one on the real line closing the set of Bethe numbers.


\section{Integral Representation for the Determinant Formula} \label{Detplusaux}
To calculate scalar-valued expectation values of local operators a nice combinatorial result for the Bethe-eigenvectors $\prod_{b=1}^M\Bs(\lambda_b)|0\rangle$ of the open $\XXZ$ chain applies. The key element is the inversion formula 
\begin{equation}
{\E_m}_{\alpha}^{\phantom{x}\beta} = \Big[\prod_{j=1}^{m-1} \big(A(s_j)+D(s_j)\big)\Big] T^\beta_{\phantom{x}\alpha}(s_m)\Big[\prod_{j=1}^m \big(A(s_j)+D(s_j)\big)^{-1}\Big]
\end{equation}
for the standard basis 
$({\E}_{\alpha}^{\phantom{x}\beta})^{\alpha^\prime}_{\phantom{x}\beta^\prime} =
\delta^{\alpha^\prime}_\alpha \delta^{\beta}_{\beta^\prime}$ at site $m$ from \cite{KMT99a}. Because it is written in terms of the monodromy $T(\lambda)$ its action on a Bethe state with \eqref{Bop} can be computed by YBA. Then some Bethe numbers $\lambda_j$ are replaced \cite{KKMNST07b} by pairwise distinct lattice inhomogeneities $\lain_k=\eta/2+s_k$ to regularize the expressions
\begin{equation}
({\E_m}_{\alpha}^{\phantom{x}\beta}) \Big[\prod_{b=1}^M\Bs(\lambda_b)\Big]|0\rangle = \sum_{\alpha_m}C_{\alpha_m}\big(\{\lambda_j\}_{j=1}^M,\{\lain_k\}_{k=1}^m\big)\Big[\prod_{b\in\alpha_m}\Bs(\mu_b)\Big]|0\rangle \quad .
\end{equation}
Here we have $\{\mu_b\}=\{\lambda_j\}_{j=1}^M \cup \{\lain_k\}_{k=1}^m$ and the summation is taken over certain subsets $\alpha_m$ of $\{1,2,\ldots,M+m\}$. For a local operator at site $m$ only the first $m$ inhomogeneities $s_1,\ldots,s_m$ enter and
their shift of $\eta/2$, $\lain_k=s_k+\eta/2$, is due to the explicit decomposition\footnote{
Note that the operator $\Bs(\lambda)$ here, \eqref{Bop}, and the corresponding expression in  Kitanine \etal\ differs by an overall prefactor and a shift of $\eta/2$ in the periodic chain operators. Looking up \cite{KKMNST07b} we find
\begin{equation*} 
\Bs_{\textrm{Kitanine}}(\lambda)= {(-1)^L}
\frac{\sh(2\lambda+\eta)}{\sh(2\lambda)}
\Big[\sh(\lambda-\eta/2+\boundary^+)B(\lambda)D(-\lambda) +
\sh(\lambda+\eta/2-\boundary^+)B(-\lambda)D(\lambda)\Big] \quad .
\end{equation*}
To make use of the normalized scalar product formula \eqref{normscalardet} one should always bear in mind, that the right Bethe vector containing $\Bs(\mu_k)$ gets its arguments from commutations starting with $\Bs(\lambda_j)$ such that the prefactors in front of the square brackets cancel due to the normalization.
}
of $\Bs(\lambda)$ in terms of the periodic chain operators. The coefficients $C_{\alpha_m}$ can be computed generically and for an illustrating example to this formula see \eqref{actiononBethe}.
\begin{propositionKKMNSTb*}
For a set of pairwise distinct numbers $\{\mu_k\}_{k=1}^M$ and Bethe roots $\{\lambda_l\}_{l=1}^M$ solving the Bethe ansatz equations (\ref{BAeq}) the normalized determinant formula for scalar products including members of the right reflection algebra reads
\begin{multline} \label{normscalardet}
\frac{\langle 0 |\!|\big[\prod_{a=1}^M\Cs(\lambda_a)\big]\big[\prod_{b=1}^M\Bs(\mu_b)\big]|0\rangle}
{\langle 0 |\!|\big[\prod_{a=1}^M\Cs(\lambda_a)\big]\big[\prod_{b=1}^M\Bs(\lambda_b)\big]|0\rangle}= \\[.2\baselineskip]
=\bigg[\prod_{a < b}\frac{\sh(\lambda_{ab})\sh(\overline{\lambda_{ab}})}
{\sh(\mu_{ab})\sh(\overline{\mu_{ab} })}\bigg]
\bigg[\prod_{l=1}^M\frac{\sh(2\mu_l+\eta)}{\sh(2\lambda_l+\eta)}\frac{\sh(2\lambda_l)}{\sh(2\mu_l)}\bigg]
\frac{\Det \big[ H(\lambda_j,\mu_k)\big]_{j,k=1,\ldots,M}}
{\Det \big[ H(\lambda_j,\lambda_k)\big]_{j,k=1,\ldots,M}}
\end{multline}
with the entry
\begin{equation}
H(\lambda_j,\mu_k) := \frac{y_j(\mu_k)-y_j(-\mu_k)}{\sh(\lambda_j-\mu_k)\sh(\lambda_j+\mu_k)}
\end{equation}
of the determinant, the shorthands $\lambda_{ab}:=\lambda_a-\lambda_b$, $\overline{\lambda_{ab}}:=\lambda_a+\lambda_b$ and the set $\{\lambda\}$ of Bethe roots included in the functions
\begin{gather}
y_j(z) = \frac{\widehat{y}(z,\{\lambda\})}{\sh(z-\lambda_j-\eta)\sh(z+\lambda_j-\eta)}\\
\widehat{y}(z,\{\lambda\}):=a(z-\eta/2)d(-z-\eta/2){\sh(z+\boundary^+ - \eta/2)\sh(z+\boundary^- - \eta/2)} \notag
\\
\times\Big[\prod_{l=1}^M\sh(z-\lambda_l-\eta)\sh(z+\lambda_l-\eta)\Big] \quad .
\end{gather}
Here $a(\lambda)=\prod_{l=1}^L\sh(\lambda-s_l+\eta)$ and $d(\lambda)=\prod_{l=1}^L\sh(\lambda-s_l)$ are the vacuum expectation values of the operators $A(\lambda)$ and $D(\lambda)$ of the periodic chain with inhomogeneities $s_l$ approaching zero in the homogeneous limit.
\end{propositionKKMNSTb*}

The Bethe ansatz equations (\ref{BAeq}) follow from
\begin{equation}
\frac{y_j(-z)}{y_j(z)} = \fraka(z)\frac{\sh(2z+\eta)}{\sh(2z-\eta)}
\frac{\sh(z+\lambda_j-\eta)\sh(z-\lambda_j-\eta)}
{\sh(z+\lambda_j+\eta)\sh(z-\lambda_j+\eta)}
\end{equation}
and can be rewritten $y_j(\lambda_j)=y_j(-\lambda_j)$, $j=1,\ldots,M$ allowing to recast the entries of the determinant in the form
\begin{multline}
H(\lambda_j,\mu_k) = \frac{y_j(\mu_k)\sh(\mu_k+\lambda_j-\eta)\sh(\mu_k-\lambda_j-\eta)}
{\sh(2\mu_k-\eta)\sh(\lambda_j-\mu_k)\sh(\lambda_j+\mu_k)} \bigg\{ 
\frac{\sh(2\mu_k-\eta)}{\sh(\mu_k+\lambda_j-\eta)\sh(\mu_k-\lambda_j-\eta)} \\-
\fraka(\mu_k)\frac{\sh(2\mu_k+\eta)}{\sh(\mu_k+\lambda_j+\eta)\sh(\mu_k-\lambda_j+\eta)}\bigg\} \quad .
\end{multline}

Considering the limit $\mu_k\to\lambda_k$ in the above expression to get in contact with the matrix elements to be calculated yields
\begin{align} \notag
\lim_{\mu_k\to\lambda_k} \frac{1}{\sh(\lambda_j+\mu_k)\sh(\lambda_j-\mu_k)}&\bigg\{ 
\frac{\sh(2\mu_k-\eta)}{\sh(\mu_k+\lambda_j-\eta)\sh(\mu_k-\lambda_j-\eta)} \\\notag
&\hspace{3em}-\fraka(\mu_k)\frac{\sh(2\mu_k+\eta)}{\sh(\mu_k+\lambda_j+\eta)\sh(\mu_k-\lambda_j+\eta)}\bigg\} \\[1ex]
=\frac{1}{\sh\eta\sh(2\lambda_j)}&\bigg[\I K_\eta(\lambda_j+\lambda_k) - \I K_\eta(\lambda_j-\lambda_k) - 
\delta^j_k \frac{\partial\ln\fraka}{\partial z}(\lambda_j)\bigg] 
\end{align}
where one separately has to treat the case $\mu_k\to\lambda_j$ accounting for the Kronecker $\delta^j_k$ by virtue of l'Hospitals rule. $K_\eta$ is the kernel from the auxiliary function,
\begin{equation}
K_\eta(\lambda)= \frac{1}{\I}\frac{\sh(2\eta)}{\sh(\lambda+\eta)\sh(\lambda-\eta)} \quad .
\end{equation}

Obviously all normalized expectation values \eqref{normscalardet} contain the elementary ratio
\begin{equation} \label{elementratio}
\frac{\Det \big[\psi(\lambda_a,\mu_b)\big]_{a,b=1,\ldots,M}}
{\Det \big[\phi(\lambda_j,\lambda_k)\big]_{j,k=1,\ldots,M}} 
= \Det \Big[\phi^{-1}(\lambda_j,\lambda_k) \psi(\lambda_k,\mu_l)\Big]_{j,l=1,\ldots,M}
\end{equation}
where on the RHS $\phi^{-1}(\lambda_j,\lambda_k)$ denote the entries of the inverse matrix and summation over $k$ is understood. Now for reshaping the RHS we closely follow \cite{GoKlSe04} as similar was done for the periodic $\XXZ$ chain. Here the entries are
\begin{align}
\phi(\lambda_j,\lambda_k) &= \frac{1}{\sh(2\lambda_j)}\bigg[\I K_\eta(\lambda_j+\lambda_k) - \I K_\eta(\lambda_j-\lambda_k) - 
\delta^j_k \frac{\partial\ln\fraka}{\partial z}(\lambda_j)\bigg] \\[1.3ex]
\psi(\lambda_j,\mu_k) &= \frac{\sh\eta\sh(2\mu_k-\eta)}
{\sh(\lambda_j-\mu_k)\sh(\mu_k-\lambda_j-\eta)\sh(\mu_k+\lambda_j-\eta)\sh(\lambda_j+\mu_k)}\notag\\ &\phantom{xxx}- \fraka(\mu_k)\frac{\sh\eta\sh(2\mu_k+\eta)}
{\sh(\lambda_j-\mu_k)\sh(\mu_k-\lambda_j+\eta)\sh(\mu_k+\lambda_j+\eta)\sh(\lambda_j+\mu_k)}
\end{align}
defining the new matrix $J(\lambda_j,\mu_l):=\phi^{-1}(\lambda_j,\lambda_k) \psi(\lambda_k,\mu_l)$.
Expressing the determinant \eqref{elementratio} in terms of a density function is presented in \mbox{appendix \ref{secdensfunc}} and summarized in the following 
\begin{lemma}
For simplicity assume $\{\mu_l\}_{l=1}^M$ to be a copy of the Bethe numbers where the first $n$ roots $\lambda_1,\ldots,\lambda_n$ are replaced by some $c$-numbers $\nu_1,\ldots,\nu_n$ thought of lattice inhomogeneities $\lain_j=\eta/2+s_j$ from the strip $|\Im(\lain_j-\eta/2)|<\varepsilon$. Then the considered determinant reduces to
\begin{equation} \label{detformulaG1}
\frac{\Det \big[\psi(\lambda_a,\mu_b)\big]_{a,b=1,\ldots,M}}
{\Det \big[\phi(\lambda_j,\lambda_k)\big]_{j,k=1,\ldots,M}} 
= \Det \big[ J(\lambda_j,\nu_l) \big]_{j,l=1,\ldots,n} = 
\Det \Big[ \frac{G(\lambda_j,\nu_l)}{\fraka^\prime(\lambda_j)} \Big]_{j,l=1,\ldots,n} 
\end{equation}
where $^\prime$ denotes a derivative and $G(\lambda,\nu)$ is the solution to the linear integral equation
\begin{multline} \label{densfunc}
G(\lambda,\nu) = \frac{\sh\eta}{\sh(\lambda+\nu)\sh(\lambda+\nu-\eta)} -
\frac{\sh\eta}{\sh(\lambda-\nu)\sh(\lambda-\nu+\eta)} \\[.2\baselineskip]+
\int_{\mathcal{C^\prime}}\frac{\D\omega}{2\pi\I}\frac{\sh(2\eta)}{\sh(\lambda-\omega+\eta)\sh(\lambda-\omega-\eta)}\frac{G(\omega,\nu)}{1+\fraka(\omega)}
\end{multline} 
on the contour $\mathcal{C}^\prime$. Here we already made use of $\fraka(\lain_j)=0$ and mind
the simple zeroes $G(0,\nu)=G(\lambda,\eta/2)=0$. The density function shows the symmetry $G(-\lambda,\nu)=-G(\lambda,\nu)$ with respect to the first argument $\lambda$ whereas the second argument $\nu$ here is restricted to the strip $|\Im(\nu-\eta/2)|<\varepsilon$ outside $\mathcal{C^\prime}$. The contour $\mathcal{C^\prime}$ excludes the hole-type solutions $\chi$ and depends on the parameter $\varepsilon$ as shown in figure \ref{allKont}.
\end{lemma}

To generalize the result let us introduce the disjoint union of the sets $\{\lambda\}=\{\lambda^+\}\cup\{\lambda^-\}$ and $\{\mu\}=\{\mu^+\}\cup\{\lambda^-\}$
and denote the cardinality of the partitions $\{\lambda^\pm\}$ by $|\lambda^\pm|$. Then along with the slightly modified function \cite{KKMNST07b}
\begin{align} \notag
\Ss_\sigma(\{\lambda^+\},\{\mu^+\}|\{\lambda^-\}) = 
\bigg[\prod_{a=1}^n 
\frac{\widehat{y}(\mu_a^+,\{\lambda\})\sh(2\mu_a^++\eta)}
{\sh(2\mu_a^+)\sh(2\mu_a^+-\eta)}
\frac{\sh(2\lambda_a^+)\sh(2\sigma_a^+\lambda_a^+-\eta)}
{\widehat{y}(\sigma_a^+\lambda_a^+,\{\lambda\})\sh(2\lambda_a^++\eta)}
\bigg]&\\
\times
\bigg[\prod_{a<b}\frac{\sh(\lambda_a^+ - \lambda_b^+)\sh(\lambda_a^+ + \lambda_b^+)}
{\sh(\mu_a^+ - \mu_b^+)\sh(\mu_a^+ + \mu_b^+)}\bigg] 
\bigg[\prod_{a=1}^n\prod_{b=1}^{M-n}
\frac{\sh(\lambda_a^+ - \lambda_b^-)\sh(\lambda_a^+ + \lambda_b^-)}
{\sh(\mu_a^+ - \lambda_b^-)\sh(\mu_a^+ + \lambda_b^-)}\bigg]
\end{align}
the normalized scalar product
\begin{equation}\label{scalarproductformula}
\frac{\langle 0 |\!|\big[\prod_{a=1}^M\Cs(\lambda_a)\big]\big[\prod_{b=1}^M\Bs(\mu_b)\big]|0\rangle}
{\langle 0 |\!|\big[\prod_{a=1}^M\Cs(\lambda_a)\big]\big[\prod_{b=1}^M\Bs(\lambda_b)\big]|0\rangle} = 
\Ss_\sigma(\{\lambda^+\},\{\mu^+\}|\{\lambda^-\}) \det \Big[\frac{G(\lambda_j^+,\mu_k^+)}{\fraka^\prime(\lambda_j^+)}\Big]_{j,k=1,\ldots,n}
\end{equation}
effectively reduces with $|\lambda^+|=|\mu^+|=n$ to an $n\times n$ matrix. The set $\{\sigma\}$ with $\sigma_j=\pm 1$ accounts for the symmetry of the Bethe roots which can be seen from the Bethe ansatz equations in the form 
$\widehat{y}(\lambda_j,\{\lambda\})\sh(-2\lambda_j-\eta) = \widehat{y}(-\lambda_j,\{\lambda\})\sh(2\lambda_j-\eta)$ for $j=1,\ldots,M$ leaving $\Ss_\sigma$ unchanged.


\section{Generating Function of the Magnetization} \label{genfunkmag}
For an illustrating example we shall now apply the integral representation of the scalar product formula \eqref{scalarproductformula} to a generating function of the $S^z$-magnetization. Note that we will assume the case of one hole-type solution on the real line accounting for $0<\boundary^\pm/\I<\pi/2$.
\begin{propositionKKMNSTa*}
Corresponding to one of the simplest non-trivial one-point functions in the open spin chain is the one-parameter generating function 
\begin{equation}
Q_m(\varphi) = \Big[\prod_{j=1}^m\big(A(s_j)+\E^\varphi D(s_j) \big)\Big]
\Big[\prod_{j=1}^m\big(A(s_j) + D(s_j) \big)^{-1}\Big]
\end{equation}
of the longitudinal magnetization 
\begin{equation}
\Big\langle\frac{1-\sigma_m^z}{2}\Big\rangle = 
\mathrm{D}_m \partial_\varphi \langle Q_m(\varphi)\rangle\big|_{\varphi=0} \quad .
\end{equation}
It includes a discrete derivative $\mathrm{D}_m u_m = u_m-u_{m-1}$ on the lattice and a continuous one with respect to $\varphi$. Its action on a Bethe state reads
\begin{align} \notag
Q_m(\varphi) \Big[\prod_{l=1}^M\Bs(\lambda_l)\Big]|0\rangle =& \sum_{n=0}^m \,
\sum_{|\lambda^+|=n}\, \sum_{|\lain^+|=n}
\Big[\prod_{j=1}^n\sum_{\sigma_j=\pm 1}\Big] \det\Big[M(\sigma_j^+\lambda_j^+,\lain_k^+)\Big]_{j,k=1,\ldots,n}\\ \notag
&\times\Big[\prod_{a=1}^n \sigma_a^+ \Big] \, W_-(\{\sigma^+\lambda^+\},\{\lain^+\})
\Big[\prod_{a=1}^n\frac{\frakb(\sigma_a^+\lambda_a^+)}{\frakb^\prime(\lain_a^+)}
\frac{1}{\sh(2\lain_a^+ - \eta)}\Big]\\ \label{actiononBethe}
&\times\Ss_\sigma^{-1}(\{\lambda^+\},\{\lain^+\}|\{\lambda^-\})
\Big[\prod_{a=1}^n \Bs(\lain_a)\Big]\Big[\prod_{b=1}^{M-n} \Bs(\lambda_b^-)\Big]|0\rangle
\end{align}
with the known matrix 
\begin{multline}\label{Mmatrix}
M(\lambda_j,\mu_k) = \frac{\sh\eta}{\sh(\lambda_j-\mu_k)\sh(\lambda_j-\mu_k-\eta)} \,+\\
+\frac{\E^\varphi\sh\eta}{\sh(\lambda_j-\mu_k)\sh(\lambda_j-\mu_k+\eta)}
\bigg[\prod_{l=1}^n\frac{\sh(\lambda_j-\lambda_l^+-\eta)\sh(\lambda_j-\mu_l^++\eta)}
{\sh(\lambda_j-\mu_l^+-\eta)\sh(\lambda_j-\lambda_l^+ +\eta)} \bigg]
\end{multline}
from the generating function of the $zz$-correlation and the function
\begin{equation} \label{wfunk}
\frac{W_-(\{\omega\},\{z\})}{W(\{\omega\},\{z\})} =
\bigg[\prod_{l=1}^n \frac{\sh(z_l + \boundary^- -\eta/2)}{\sh(\omega_l + \boundary^-  -\eta/2)}\bigg]
\bigg[\frac{\prod_{a,b=1}^n \sh(z_b +\omega_a - \eta)}
{\prod_{a<b}\sh(z_a+z_b-\eta)\sh(\omega_a+\omega_b-\eta)}\bigg]
\end{equation}
\begin{equation}
W(\{\omega\},\{z\}) = \bigg[\prod_{a,b=1}^n\frac{\sh(z_b-\omega_a-\eta)\sh(z_b-\omega_a+\eta)}
{\sh(\omega_a-\omega_b-\eta)\sh(z_a-z_b+\eta)}\bigg]
\end{equation}
picking out the left boundary with $\boundary^-$ to start counting the lattice sites. All inhomogeneities entering the generating function are included within the expressions \cite{GoKlSe04}
\begin{equation} \label{frakbeh}
\frakb(\lambda)=\bigg[\prod_{l=1}^m\frac{\sh(\lambda-\lain_l)}{\sh(\lambda-\lain_l+\eta)}\bigg] 
\quad , \quad 
\frac{1}{\frakb^\prime(\lain_j)} = \frac{\prod_{l=1}^m\sh(\lain_j-\lain_l+\eta)}{\prod_{\substack{l=1\\l\not=j}}^m\sh(\lain_j-\lain_l)} \quad .
\end{equation}
The function $\Ss_\sigma$ already appeared in the scalar product formula, whereas $\sigma_j=\pm1$ accounts for the symmetry of the Bethe roots.
\end{propositionKKMNSTa*}

Here the combinatorial part is expressed by the set of all ordered pairs $(\{\lambda^+\},\{\lambda^-\})$ of fixed cardinality $|\lambda^+|$ and $|\lain^+|$ respectively indexing the sums. Switching to the normalized scalar product the expectation value of the generating function can be written in terms of the density function $G(\sigma_j\lambda_j,\nu)=\sigma_jG(\lambda_j,\nu)$,
\begin{align} \notag
\langle Q_m(\varphi)\rangle =& 
\frac{\langle 0 |\!|\big[\prod_{a=1}^M\Cs(\lambda_a)\big] Q_m(\varphi)\big[\prod_{b=1}^M\Bs(\lambda_b)\big]|0\rangle}
{\langle 0 |\!|\big[\prod_{a=1}^M\Cs(\lambda_a)\big]\big[\prod_{b=1}^M\Bs(\lambda_b)\big]|0\rangle}\\
=& \sum_{n=0}^m \,\, \sum_{|\lambda^+|=n} \,\, \sum_{|\lain^+|=n} \,\,\Big[\prod_{j=1}^n\sum_{\sigma_j=\pm 1}\Big] 
\bigg[\prod_{a=1}^n\frac{\frakb(\sigma_a^+\lambda_a^+)}
{\frakb^\prime(\lain_a^+)}\frac{1}{\sh(2\lain_a^+-\eta)}\bigg]\label{pregenfunc}\\\notag
&\times W_-(\{\sigma^+\lambda^+\},\{\lain^+\}) \det\Big[M(\sigma_j^+\lambda_j^+,\lain_k^+)\Big]_{j,k=1,\ldots,n}
\det\Big[\frac{G(\sigma_j^+\lambda_j^+,\lain_k^+)}{\fraka^\prime(\lambda_j^+)}\Big]_{j,k=1,\ldots,n}
\end{align}

The last step now is to get rid of the explicit dependence on Bethe roots by integrals  according to the following
\begin{lemma} \label{lemmacontint}
Let $f(\omega_1,\ldots,\omega_n)$ be a complex function, symmetric in its arguments and equal to zero if any two of its arguments agree up to a sign. Furthermore if it is analytic on and inside the simple $n$-fold contour $(\mathcal{C^\prime})^n$ and shows a simple zero at $\omega_j=0$ to compensate the first order pole of the auxiliary function $1/(1+\fraka)$ then
\begin{equation}
\sum_{|\lambda^+|=n} \,\sum_{\sigma_1^+=\pm1}\ldots\sum_{\sigma_n^+=\pm1} \frac{f(\sigma_1^+\lambda_1^+,\ldots,\sigma_n^+\lambda_n^+)}{\prod_{l=1}^n\fraka^\prime(\lambda_l^+)} = \frac{1}{n!} \bigg[\prod_{l=1}^n\int_{\mathcal{C^\prime}}\frac{\D\omega_l}{2\pi\I}\frac{1}{1+\fraka(\omega_l)}\bigg] f(\omega_1,\ldots,\omega_n) \, .
\end{equation}
\end{lemma}

For lattice inhomogeneities from the strip $|\Im(\lain_k-\eta/2)|<\varepsilon$ all poles with respect to the variable $\omega_j$ of the density function $G(\omega_j,\lain_k)$ lie outside the contour $\mathcal{C^\prime}$.
The singularity of the function $W_-(\{\omega\},\{\lain\})$ at $\omega_j=\eta/2-\boundary^-$ is balanced by the simple zero of $1/(1+\fraka(\omega_j))$ and because of the density function $G(0,\nu)=0$ the expression in \eqref{pregenfunc} meets along with the determinant property the conditions of lemma \ref{lemmacontint}. \\

However, the same technique can be applied to the $\lain^+$-summation with inhomogeneities $\lain_k$ in the vicinity of $\eta/2$ and thus outside $\mathcal{C^\prime}$. For a function $f$ with the properties from above except the simple zero at $\omega_j=0$ the corresponding integrals read \cite{GoKlSe04}
\begin{equation} \label{residuebackwards}
\sum_{|\xi^+|=n}  \frac{f(\lain_1^+,\ldots,\lain_n^+)}{\prod_{l=1}^n\frakb^\prime(\lain_l^+)} = \frac{1}{n!} \bigg[\prod_{l=1}^n\int_{\Gamma}\frac{\D z_l}{2\pi\I}\frac{1}{\frakb(z_l)}\bigg] f(z_1,\ldots,z_n) \quad .
\end{equation}

\begin{figure}
\begin{center}
\epsfig{file=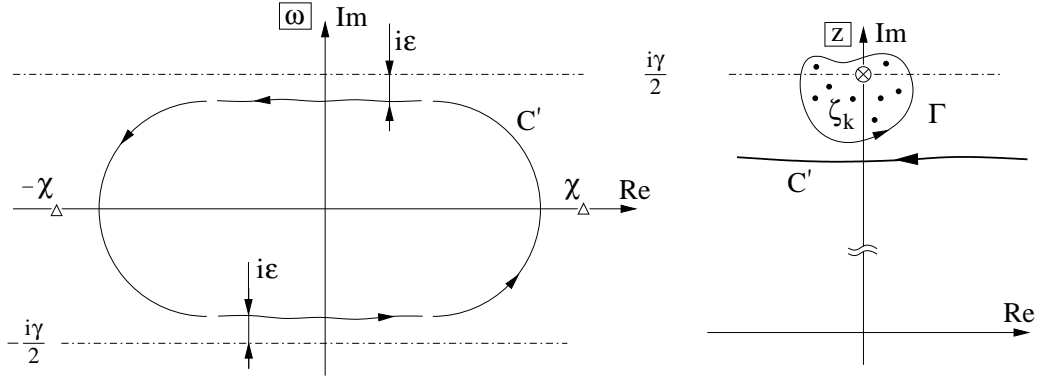,height=5cm}
\caption{\label{allKont} In the massless case $\eta=\I\gamma$, $0<\gamma<\pi/2$ the contour $\mathcal{C^\prime}$ is limited by the hole-type solution $\chi$ as depicted in the left panel for a small $\varepsilon\ll 1$. The lattice inhomogeneities $\lain_k$ lie in the vicinity of $\eta/2$ outside $\mathcal{C^\prime}$ and are counterclockwisely surrounded by $\Gamma$ (right panel).}
\end{center}
\end{figure}

With respect to the integrand the contour $\Gamma$ lies in the strip $|\Im(z-\eta/2)|<\varepsilon$  and surrounds all inhomogeneities $\lain_1,\ldots,\lain_m$. In addition due to the simple zero $G(\lambda,\eta/2)=0$ the point $\eta/2$ can be enclosed enabling the homogeneous limit $\lain_k\to\eta/2$ yielding
\begin{proposition*}
Let $m$ be a site index counted from the left boundary and consider the functions $\frakb(\lambda)$, $W_-(\{\omega\},\{z\})$, $M(\omega_j,z_k)$ and $G(\omega_j,z_k)$ according to \eqref{frakbeh}, \eqref{wfunk}, \eqref{Mmatrix} and \eqref{densfunc}. Then the multiple integral representation of the generating function reads
\begin{align} \notag
\langle Q_m(\varphi)\rangle = \sum_{n=0}^m & \frac{1}{(n!)^2}
\bigg[\prod_{l=1}^n \int_{\mathcal{C^\prime}}\frac{\D\omega_l}{2\pi\I}
\frac{\frakb(\omega_l)}{1+\fraka(\omega_l)}
\int_{\Gamma}\frac{\D z_l}{2\pi\I}\frac{1}{\frakb(z_l)}
\bigg]\, W_-(\{\omega\},\{z\})\\[.1\baselineskip]
& \times\det\Big[M(\omega_j,z_k)\Big]_{j,k=1,\ldots,n}
\det\Big[\frac{G(\omega_j,z_k)}{\sh(2z_k-\eta)}\Big]_{j,k=1,\ldots,n} \quad .
\end{align}
The contours for the massless case are depicted in figure \ref{allKont} and in the homogeneous limit the auxiliary function $\fraka(z)$ is determined from the non-linear integral equation \eqref{NLIE}. 
\end{proposition*}

\subsection*{Thermodynamic Limit}
Rewriting the density function \eqref{densfunc} as the sum 
$G(\lambda,\nu) = G^+(\lambda,\nu)-G^+(\lambda,\eta-\nu)$ the partial density $G^+$ satisfies for $|\Im (\nu^\prime-\eta/2)|<\varepsilon$
\begin{equation}
G^+(\lambda,\nu^\prime) = -\frac{\sh\eta}{\sh(\lambda-\nu^\prime)\sh(\lambda-\nu^\prime+\eta)} +\int_{\mathcal{C^\prime}}\frac{\D\omega}{2\pi\I}\frac{\sh(2\eta)}{\sh(\lambda-\omega+\eta)\sh(\lambda-\omega-\eta)}\frac{G^+(\omega,\nu^\prime)}{1+\fraka(\omega)} \quad .
\end{equation}

\begin{figure}
\begin{center}
\epsfig{file=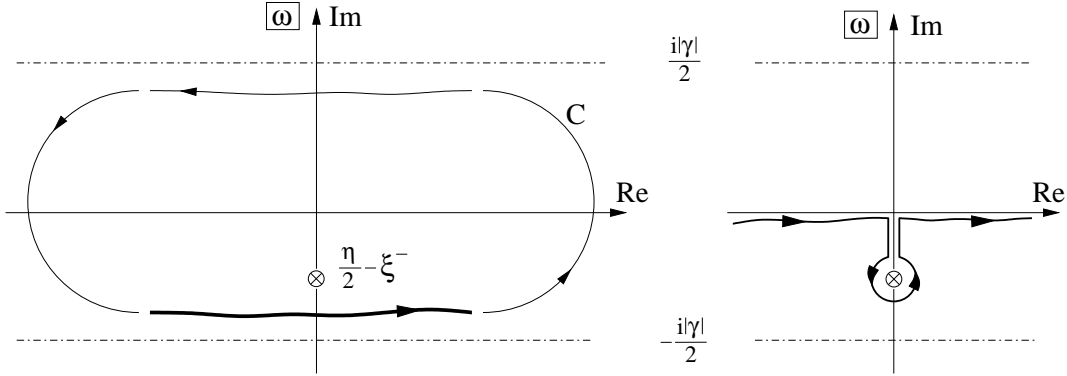,height=5cm}
\caption{\label{TLKont} In the thermodynamic limit only the lower part of the contour $\mathcal{C}$ remains (left panel). Moving it towards the real axis the poles at $\omega_j=\eta/2-\boundary^-$ of the function $W_-$ must not be crossed (right panel).}
\end{center}
\end{figure}

In the limit of infinitely many lattice sites $L\to\infty$, thus $\chi\to\infty$ and $\mathcal{C^\prime}\to\mathcal{C}$, the auxiliary function $\fraka$ is dominating for $\gamma<0$ the upper part of the contour $\mathcal{C}$, 
\begin{equation}
\ln\fraka(\lambda) \sim 2L\frac{\sh(\lambda-\I\gamma/2)}{\sh(\lambda+\I\gamma/2)} \quad ,
\end{equation}
such that only the lower part remains (figure \ref{TLKont}, left panel). Clearly, the validity range of the variable $\nu^\prime$ extends according to the pole structure of the driving term and the density
$G^+(\lambda,\nu^\prime)\to\I\pi\rho(\lambda,\nu^\prime)$ satisfies in this limit
\begin{equation}
\rho(\lambda,\nu^\prime) + \int\displaylimits_{-\infty}^\infty\frac{\D\omega}{2\pi}
\frac{\I\sh(2\eta)\rho(\omega,\nu^\prime)}{\sh(\lambda-\omega+\eta)\sh(\lambda-\omega-\eta)}= 
\frac{\I}{\pi}\frac{\sh\eta}{\sh(\lambda-\nu^\prime)\sh(\lambda-\nu^\prime+\eta)} \quad .
\end{equation}

Remember, the variable $\nu^\prime$ takes the values of $\nu$ and $\eta-\nu$ where $\nu$ is located in the vicinity of $\eta/2$.
Applying the thermodynamic limit to the generating function one is directly led to the result of Kitanine \etal \cite{KKMNST08a} where $\gamma<0$ is assumed,
\begin{align} \notag
\langle Q_m(\varphi)\rangle =\sum_{n=0}^m & \frac{1}{(n!)^2}
\bigg[\prod_{l=1}^n 
\int_{C_D}\!\!\!\!{\D\omega_l}
\int_{\Gamma}\frac{\D z_l}{2\pi\I}
\frac{\frakb(\omega_l)}{\frakb(z_l)}\bigg] \,W_-(\{\omega\},\{z\})\\[.1\baselineskip]
&  \times \det\Big[M(\omega_j,z_k)\Big]_{j,k=1,\ldots,n}
\det\Big[\frac{\rho(\omega_j,z_k)-\rho(\omega_j,\eta-z_k)}{2\sh(2z_k-\eta)}\Big]_{j,k=1,\ldots,n} \quad .
\end{align}
$C_D$ consists of the real line and an additional counterclockwisely closed contour around the pole $\omega=\eta/2-\boundary^-$ dependent on the value of the boundary parameter $\boundary^-$. The condition $-|\gamma|/2<\Im\boundary^-<0$ for this additional contribution to be included can easily be seen from \mbox{figure \ref{TLKont}}, right panel.


\section{Conclusion}
In this paper we stated for the open $\XXZ$ spin chain a non-linear integral equation for an auxiliary function accounting for the ground state of the model in regions $I$, $I\!I$, $I\!I\!I$, $IV$ and $V\!I\!I$. In the other cases a neighbouring state of zero magnetization is described 
which depends on the value of the anisotropy and boundary fields and reduces just to region $V\!I$ in the $\XXX$ limit. Our formulae are derived for the Bethe ansatz solvable case of $S^z$-conserving boundaries and are valid for a finite number of even lattice sites. In the main part we showed how to combine this method with the scalar product formula of Bethe vectors. This yielded a linear integral equation whose solution builds on the well known determinant representing scalar products. As an example we derived for a certain generating function of the $S^z$-magnetization a multiple integral representation showing the correct thermodynamic limit and matching the result from exact diagonalization for small lattice sites.

For the derivation we assumed a distribution of Bethe solutions having one hole-type solution on the real line. Thus we had to choose a closed contour $\mathcal{C^\prime}$ for the integral representations of the determinant formula and generating function differing from the canonical contour $\mathcal{C}$ of the auxiliary function. But this is already the general case. In the simplest issue of field parameters
$0 > \boundary^\pm/\I > -\pi/2$ and $\pi/2 \geq \boundary^\mp/\I > 0$ (regions $V\!I\!I$, $V\!I\!I\!I$ and $I\!X$) no \lq holes\rq\ have to be taken into account such that the canonical contour even applies for the integral representations. Therefore we expect this to be a good starting point for numerical considerations. Interesting as well would be a refinement of the regions $V$, $V\!I$, $V\!I\!I\!I$ and $I\!X$ for the Bethe ansatz with reference to the alternation in the description between the ground state and its zero-magnetization neighboured state which both degenerate in the thermodynamic limit.

\begin{figure}
\begin{center}
\epsfig{file=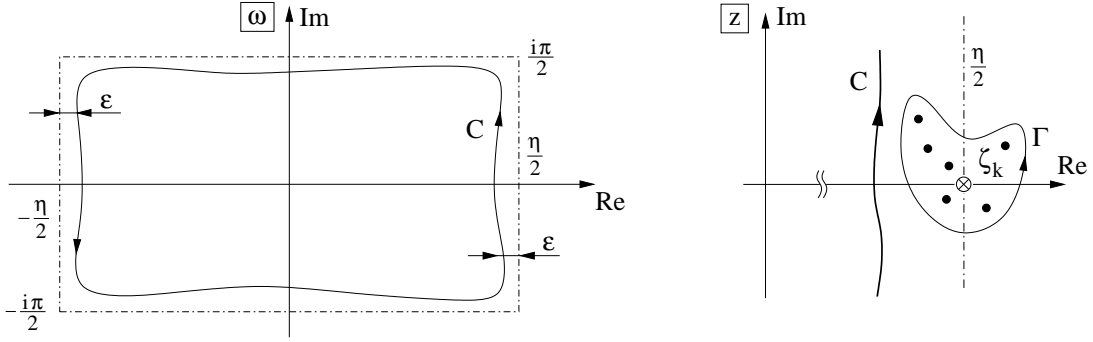,height=4.5cm}
\caption{\label{masscont}  In the massive case $\eta>0$ the canonical contour $\mathcal{C}$ with a small $\varepsilon\ll1$ is depicted in the left panel. The lattice inhomogeneities $\lain_k$ lie in the vicinity of $\eta/2$ outside $\mathcal{C}$ and are counterclockwisely surrounded by the closed contour $\Gamma$ (right panel).}
\end{center}
\end{figure}

In this article we restricted our derivation to the massless case $\eta=\I\gamma$ with $0<\gamma<\pi/2$. For the region $\pi/2<\gamma<\pi$ the imaginary parts of the contours take values $\pm(\gamma-\pi/2-\varepsilon)$, $\varepsilon\ll1$
to ensure the hole-type solutions to be located outside the contour due to $\I\pi$-periodicity. We would like to note that the results can be extended to the massive regime of real crossing parameters $\eta>0$, where the contours have to be redefined according to figure \ref{masscont}.\\

{\bf Acknowledgements.}  
The authors would like to thank J. Damerau and H. Frahm for helpful discussions. AS and TW gratefully acknowledge financial support by the Deutsche Forschungsgemeinschaft under grant numbers \mbox{SE~1742/1-2} and FR~737/6 .


\begin{appendix}
\section{The Density Function}\label{secdensfunc}

In order not to overload the notation lets consider the set 
$\{\mu_l\}_{l=1}^M = \{\nu_j\}_{j=1}^n\cup\{\lambda_l\}_{l=n+1}^M$. Then the expression under the determinant of the RHS of (\ref{elementratio}) reads in components
\begin{equation}
\psi(\lambda_j,\nu_l) = \frac{\I}{\sh(2\lambda_j)} 
\sum_{k=1}^M \Big[ K_\eta(\lambda_j+\lambda_k) -K_\eta(\lambda_j-\lambda_k) \Big] J(\lambda_k,\nu_l) -\frac{J(\lambda_j,\nu_l)}{\sh(2\lambda_j)}
\frac{\partial\ln\fraka}{\partial z}(\lambda_j) \quad ,
\end{equation}
solved for the part containing the logarithmic derivative of the auxiliary function
\begin{align} \label{defJ}
{J(\lambda_j,\nu_l)}&
\frac{\partial\ln\fraka}{\partial z}(\lambda_j) = {\I} 
\sum_{k=1}^M \Big[ K_\eta(\lambda_j+\lambda_k) - K_\eta(\lambda_j-\lambda_k) \Big]J(\lambda_k,\nu_l)\notag\\
&+ \Big[\frac{\sh\eta}{\sh(\nu_l-\lambda_j)\sh(\nu_l-\lambda_j-\eta)} - \frac{\sh\eta}{\sh(\nu_l+\lambda_j)\sh(\nu_l+\lambda_j-\eta)}\Big]\notag\\
&- \Big[\frac{\sh\eta}{\sh(\nu_l-\lambda_j)\sh(\nu_l-\lambda_j+\eta)} - \frac{\sh\eta}{\sh(\nu_l+\lambda_j)\sh(\nu_l+\lambda_j+\eta)}\Big]\fraka(\nu_l) \quad .
\end{align}

Strikingly the definition of $J(\lambda,\nu)$ according to (\ref{defJ}) is compatible with the constraint $J(\lambda,\nu) = -J(-\lambda,\nu)$ and the properties $\fraka(z)=1/\fraka(-z)$, $\fraka^\prime(\lambda_j)=\fraka^\prime(-\lambda_j)$ of the auxiliary function. Considering
\begin{equation}
F(\lambda_j,\nu_l) := {J(\lambda_j,\nu_l)}
\frac{\partial\ln\fraka}{\partial z}(\lambda_j)
\end{equation}
for arbitrary arguments $\lambda_j$ the analytic properties $\res_{\lambda=\pm\nu}F(\lambda,\nu) = 1 + \fraka(\nu)$ beside the single zero $F(0,\nu)=0$ are known from the RHS of (\ref{defJ}) such that
\begin{multline}
\int_{\mathcal{C^\prime}}\frac{\D\omega}{2\pi\I}\frac{\sh(2\eta)}{\sh(\lambda-\omega+\eta)\sh(\lambda-\omega-\eta)}\frac{F(\omega,\nu)}{1+\fraka(\omega)} = F(\lambda,\nu) \\
+(1+\fraka(\nu))\Big[\frac{\sh\eta}{\sh(\lambda+\nu)\sh(\lambda+\nu-\eta)}+
\frac{\sh\eta}{\sh(\lambda-\nu)\sh(\lambda-\nu-\eta)}\Big]
\end{multline}
holds where all Bethe roots $\lambda_j$ and the variable $\nu$ except the hole-type solution $\chi$ on the real line are supposed to lie inside $\mathcal{C^\prime}$. Redefining
$F(\lambda,\nu):=-(1+\fraka(\nu))G(\lambda,\nu)$
we are led to
\begin{multline}\label{densfunc2}
G(\lambda,\nu) = \frac{\sh\eta}{\sh(\lambda-\nu)\sh(\lambda-\nu-\eta)} +
\frac{\sh\eta}{\sh(\lambda+\nu)\sh(\lambda+\nu-\eta)} \\[.2\baselineskip]+
\int_{\mathcal{C^\prime}}\frac{\D\omega}{2\pi\I}\frac{\sh(2\eta)}{\sh(\lambda-\omega+\eta)\sh(\lambda-\omega-\eta)}\frac{G(\omega,\nu)}{1+\fraka(\omega)} \quad .
\end{multline} 

Unfortunately in our case $\nu$ should be a lattice inhomogeneity $\lain_j$ taken from the strip $|\Im(\nu-\eta/2)|<\varepsilon$. Thus $G(\lambda,\nu)$ is due to an additional residue and $\fraka(\lain_j)=0$ the solution to the linear integral equation \eqref{densfunc}. The considered determinant (\ref{elementratio}) is calculated from
\begin{equation} \label{detformulaG}
\frac{\Det \big[\psi(\lambda_a,\mu_b)\big]_{a,b=1,\ldots,M}}
{\Det \big[\phi(\lambda_j,\lambda_k)\big]_{j,k=1,\ldots,M}} 
= \Det \big[ J(\lambda_j,\nu_l) \big]_{j,l=1,\ldots,n} = 
\Det \Big[ \frac{(1+\fraka(\nu_l))G(\lambda_j,\nu_l)}{\fraka^\prime(\lambda_j)} \Big]_{j,l=1,\ldots,n} 
\end{equation}
and making again use of $\fraka(\lain_j)=0$ for the variable $\nu_l=\lain_j=\eta/2+s_j$ reduces the expression \eqref{detformulaG} to \eqref{detformulaG1}. Clearly, explicitly using $\fraka(\lain_j)=0$ changes $G(\lambda,\nu)$ for the argument $\nu$ away from $\lain_j$ compared to the original definition. But because $G(\lambda,\nu)$ is only used in combination with simple poles and the residue theorem (c.f.\ \eqref{residuebackwards}) just its unchanged value $G(\lambda,\lain_j)$ at the lattice inhomogeneity $\lain_j$ is relevant.

\end{appendix}



\providecommand{\bysame}{\leavevmode\hbox to3em{\hrulefill}\thinspace}
\providecommand{\MR}{\relax\ifhmode\unskip\space\fi MR }
\providecommand{\MRhref}[2]{%
  \href{http://www.ams.org/mathscinet-getitem?mr=#1}{#2}
}
\providecommand{\href}[2]{#2}

\end{document}